\documentclass[a4paper,fleqn]{cas-dc}

\usepackage[numbers]{natbib}

\usepackage{color}

\usepackage[switch]{lineno}

\usepackage{multirow}
\usepackage{makecell}
\usepackage{subcaption}
\usepackage{adjustbox}
\usepackage{bbold}
\usepackage{float}
\usepackage{graphicx}

\usepackage{xcolor}
\usepackage{hyperref}
\usepackage{tabularx}

\hypersetup{
    colorlinks=true,
    linkcolor=black,
    filecolor=magenta,      
    urlcolor=blue,
    }

\begin{document}
\let\WriteBookmarks\relax
\def\floatpagepagefraction{1}
\def\textpagefraction{.001}

\shorttitle{A unified platform for GIS data management and analytics}

\shortauthors{Piccoli et~al.}

\title [mode = title]{A unified platform for GIS data management and analytics}

\address[1]{Istituto Nazionale di Fisica Nucleare,Milano,Italy}

\address[2]{University of Milano-Bicocca,Milano,Italy}

\author[1]{Flavio Piccoli}[type=editor,
                        orcid=0000-0001-7432-4284]
\ead{flavio.piccoli@unimib.it}
\cormark[1]
\cortext[cor1]{Corresponding author}

\author[2]{Simone Locatelli}[type=editor,
                        orcid=0000-0003-3929-8409]
\ead{s.locatelli29@campus.unimib.it}
                        
\author[2]{Paolo Napoletano}[type=editor,
                        orcid=0000-0001-9112-0574]
\ead{paolo.napoletano@unimib.it}

\author[2]{Raimondo Schettini}[type=editor,
                        orcid=0000-0001-7461-1451]
\ead{raimondo.schettini@unimib.it}

\begin{abstract}
In this work we propose a software platform for the collection, visualization, management and  analysis of heterogeneous and multisource data for soil characteristics estimation. The platform is designed in such a way that it can easily handle proximity, airborne and spaceborne data and provides to the final user all the tools and proper visualizations to perform precision agriculture. The proposed software allows an easy integration of new data (that can be performed directly on-board of the acquisition device) and the possibility to add custom predictive systems for soil characteristic estimation. Usability experiments show that the platform is easy to use and effective.

\end{abstract}


\maketitle

\section{Introduction}
\label{sec:introduction}
With the increase of population and the decrease of resources, sustainability is becoming an increasingly important issue and technological innovation can help to optimize all those processes that waste a lot of resources \cite{velten2015sustainable}.

Agriculture is one of those sectors. In this context, the adoption of precision agriculture (PA) can dramatically contribute to improve sustainability~\cite{pierce1999aspects}. 
Quoting the The International Society of Precision Agriculture\footnote{\url{https://www.springer.com/journal/11119}}, ``\emph{PA is a management strategy that gathers, processes and analyzes temporal, spatial and individual data and combines it with other information to support management decisions according to estimated variability for improved resource use efficiency, productivity, quality, profitability and sustainability of agricultural production}''.

Data-driven Artificial Intelligence (AI) methods can help to support and/or automate management decisions~\cite{lindblom2017promoting}. Agricultural Decision Support Systems (DSS) have been created especially to support farmers in taking effective 
 on-going and/or future decisions ~\cite{lindblom2017promoting,shafi2019precision}.

DSS will guide effective agronomic practices to optimize the phases of mechanical processing, fertilization, weeding, irrigation, distribution of pesticides and harvesting to a very high degree of detail, by dosing the intensity according to the characteristics of the soil and the crop that have been measured and loaded into the system~\cite{ayaz2019internet}.

This significantly improves the entire process, thus reducing waste to a minimum and optimizing each intervention \cite{sishodia2020applications}.
The use of data-driven technologies however require an expensive process of data collection and analysis.
With the aim of avoiding mistakes, this process must follow strict acquisitions protocols and should be supported by an ad-hoc software platform \cite{tantalaki2019data, linaza2021data}.

Furthermore, the scalability of the DSS on new areas (for which no samples are available) requires the use of AI-based predictive systems which are capable of performing an estimation of the soil characteristics from remote sensors~\cite{mulla2021satellite}.
Specifically, it is possible to perform soil characteristics estimation from airborne sensors such as drones or aerostatic baloons and/or spaceborn sensors such as satellites~\cite{elijah2018overview}. 

Data-driven AI methods require large amount of data to be stored, managed, annotated, visualized, explored, and eventually edited~\cite{strickland2022andrew}. Most importantly, the DSSs should provide an intuitive and effective grahical interface to permit result interpretability for human supervision and understanding~\cite{kukar2019agrodss,coppola2019landsupport}

In this work we propose the Pignoletto web-platform, which addresses all the aforementioned topics for supporting data-driven precision agriculture. In particular, the proposed platform:

\begin{itemize}
    \item supports data collection and management from heterogeneous data sources
    \item provides all the tools for data visualization and decision-making
    \item provides soil characteristics AI-based predictors
    \item provides a procedure to include custom predictors
\end{itemize}

The platform has been populated with geophysical fields measurements acquired in the region of Lombardy (Italy) combined with airborne sensors (ionizing gamma radiation, optical hyperspectral, thermal multispectral), satellite information (PRISMA and Copernicus) and soil proximity measures. A great portion of the aforementioned data have been acquired within the Pignoletto Project~\footnote{(more info here \url{https://www.pignolettomibinfn.it/})}.

The article is organized as follows: in section \ref{sec:context} we describe the technological context, in section \ref{sec:architecture} we describe in detail its architecture, in section \ref{sec:deployment} we present the hardware and software configurations used for the development of the application, in section \ref{sec:usability} we present the results obtained through a survey on the usability of the platform
and finally in section \ref{sec:conclusion} we present our conclusions.

\section{Scientific and technological context}
\label{sec:context}

Geographic Information Systems (GIS) has emerged as an effective tool for the macro and micro level mapping of natural resources and artificial deployments, thus, their use extends to different sectors, from research to business~\cite{longley2005geographic}.
There are several GIS software on the market but the most widely used are QGIS \cite{QGIS} and ArcGIS~\cite{booth2001getting}.

GIS are a powerful set of tools for collecting, storing and retrieving, transforming and visualizing spatial data.
Their ability to analyze and visualize agricultural environments and workflows has proven to be beneficial for the agricultural sector \cite{sebastianelli2021automatic}.

GIS technology is becoming an essential tool for combining different sources of data, such as data acquired by drones, airborne sensors and satellites~\cite{ma2001mapping}.

In the literature, GIS technology has been used in various scenarios. Frigerio et al. \cite{frigerio2016gis} proposed a methodology that exploits socio-economic factors to identify how different areas of the country can react to catastrophic natural events.

Blanco et al. \cite{blanco2018agricultural}, instead, exploited the potential of GIS systems to monitor the use and waste of plastic in the Puglia (Italy) region, in particular the tool developed allows stakeholders to quantify and localize areas characterized by intensive plastic production, thus allowing the study of the most suitable areas for collection centers.

Mancini et al. proposed a methodology for the management of landslide hazard \cite{mancini2010gis}, while Ladisa et al. proposed a methodology for the evaluation of desertification risks related to specific areas as in \cite{ladisa2012gis}.

GIS tools are also playing an increasingly important role in precision agricultural, thus helping farmers to increase production, reduce costs and manage their land more efficiently~\cite{michelon2019software, bazzi2019agdatabox}. 

Furthermore, GIS have proven to be valuable tools in preventing soil erosion and degradation processes~\cite{petito2022impact}.  

\section{Pignoletto use case}
In this context, the Pignoletto platform aims to target the specific agricultural needs of the Lombardy region which are the automatic creation of soil characteristics maps of the agricultural lands on basis of proximity, remote and hybrid chemical and spectral measures. The platform and its components are open-source tools and it leaves the possibility of being easily modifiable and adaptable for other locations or specific needs. This will support local farmers to adapt their practice in crop management~\footnote{\url{https://www.pignolettomibinfn.it/il-progetto}}.

The project considers two different types of land measures: 1) laboratory $\mathcal{L}$ measures and the drone acquisitions $\mathcal{D}$. 
The former set is collected in a controlled environment to support the training of a deep-learning-based system $M$ able to estimate soil properties $p$ from the spectral information $h$, i.e. $M(h)=\hat{p}$. To this aim, each sample $(h,p) \in \mathcal{L}$ is composed by the hyperspectral signal $h$ together with the soil properties $p$ measured by an expert. The latter set is composed by drone acquisitions that will be loaded at test time. These acquisitions are acquired in the wild. Once they are loaded in the system, one or more predictive systems $M$ will perform an estimation of the soil properties $\hat{p}$ associated to this signal.

\begin{figure*}[tb] %
\centering
\begin{adjustbox}{width=0.8\linewidth}
\includegraphics[width=\columnwidth]{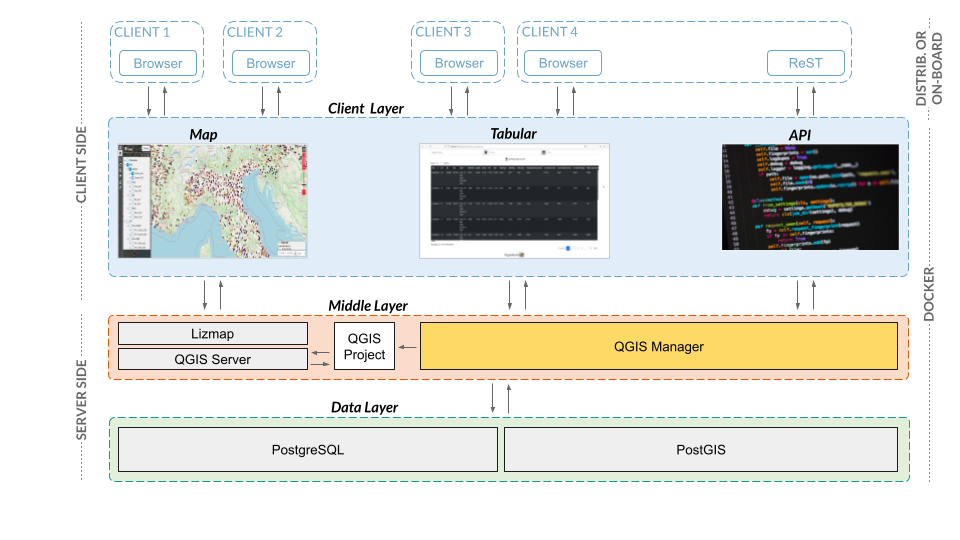}
\end{adjustbox}
\caption{Architecture of the software. Clients can interact with the system through the \emph{Map}, the \emph{Tabular} or the RESTful API. Our class QGISManager furnishes all the functions to perform CRUD operations on the platform database.}
\label{fig:architecture}
\end{figure*}

\section{Software architecture}
\label{sec:architecture}

The proposed platform, as also discussed in section 1, is designed to handle geophysical fields measurements acquired in the region of Lombardy (Italy)
combined with airborne sensors (ionizing gamma radiation, optical hyperspectral, thermal multispectral), satellite information (PRISMA and Copernicus) and soil proximity measures.

Figure \ref{fig:architecture} shows all components of proposed Pignoletto platform and their interactions. The  platform components can be grouped into \emph{server} and \emph{client-side} elements. The \emph{client-side} is composed of several tools for interacting with data. The  \emph{Map} tool allows visual-based interaction with soil maps through a web browser; the \emph{Tabular} tool allows the interaction with data in a tabular format; and the \emph{API} (Application Programming Interface) tool allows the interaction with the platform through a computer program. The three interaction modalities are described more in detail respectively in subsections \ref{sub:map}, \ref{sub:backend} and \ref{sub:api}. 

The \emph{server-side} is composed of a \emph{middle layer} which orchestrates the interactions between the clients and the \emph{data layer}. Specifically, for the \emph{Map} component, the \emph{middle layer} creates, at run time, a QGIS project and the corresponding Lizmap~\cite{laurent2018online} configuration for the creation of the web interface. For the \emph{Tabular} and \emph{API} components, the middle layer exposes proper functions for data uploading, querying and editing.  Data is stored in a PostgreSQL database with a CRUD-type storage engine (create, read, update and delete). Please refer to subsection \ref{sub:db} for further details.

The use of QGIS and its utilities allows the sharing of platform geographical data through Web Map Service (WMS) and Web Feature Service (WFS) which are well known standards in the GIS community~\cite{michaelis2008web}.

All technologies used for the creation of the platform are free and open source. To make the system usable in any context and to maximize reproducibility, a Docker~\cite{anderson2015docker} environment containing all the necessary software and data samples was created. Here you can find the GitHub repository of the proposed platform~\footnote{\url{https://github.com/SimoLoca/Pignoletto\_platform}}.

The platform considers different types of users: 1) anonymous clients, 2) registered clients, 3) administrator. Anonymous users can only navigate the \emph{Map} component without any possibilities to edit the content. Registered users can interact with all the components of the \emph{client layer}. They can add new data to the platform but they can not delete existing data. Newly subscribed users must be manually approved by an administrator. The administrator has all the privileges to edit platform data.

\begin{figure*}[tb]
    \centering
    \begin{tabular}{ccc}
         \includegraphics[width=0.28\textwidth]{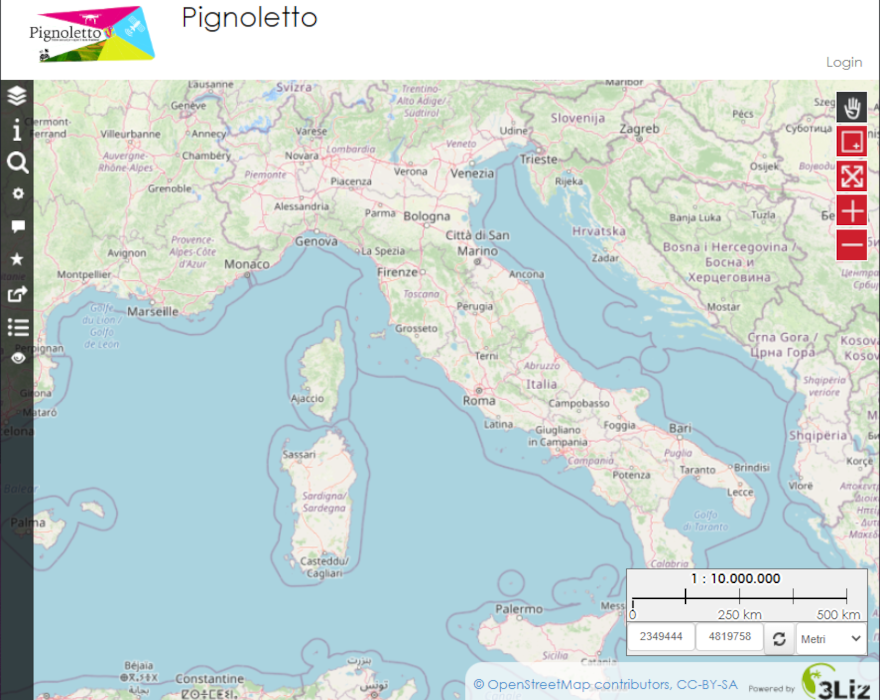}&  \includegraphics[width=0.28\textwidth]{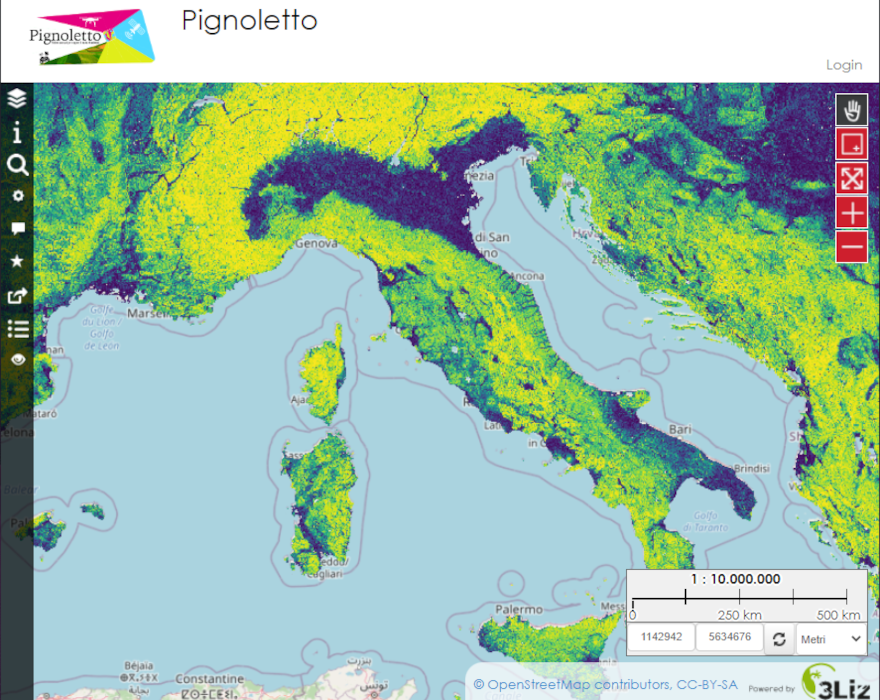} & 
        \includegraphics[width=0.28\textwidth]{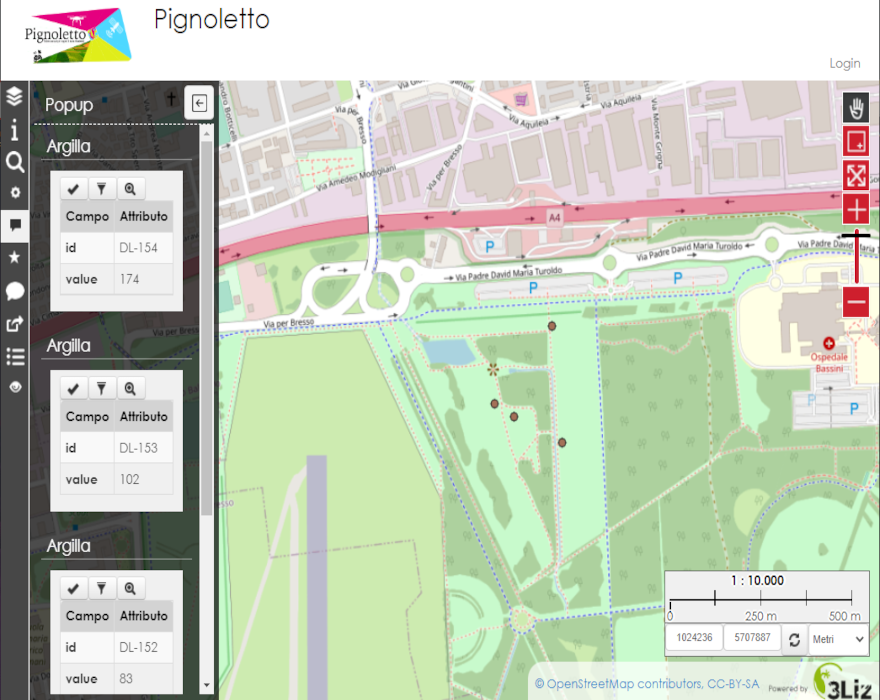}\\
         (a)&(b)&(c)\\
    \end{tabular}
    \caption{Main functionalities of the platform. (a) Initial view (b) Raster layer selected (c) Zoom on acquisitions}
    \label{fig:demo}
\end{figure*}
\subsection{Map}
\label{sub:map}

The \emph{Map} is displayed as a Web Map Service application built using a custom version of the Lizmap API \footnote{\url{https://github.com/3liz/lizmap-plugin}}~\cite{laurent2018online}.
It is built on top of a QGIS server that displays the information contained in a QGIS project on a geographical map. It includes several Lizmap tools that allow the user to interact and explore the data. Table \ref{tab:lizmap-utilities} describes these tools in detail. For instance, it is possible to filter data on the basis of their properties, to show the information about the collected data in a tabular way, etc. 

The QGIS project rendered through the \emph{Map} is created by the \emph{middle layer} when the platform is started. Contextually, the \emph{middle layer} creates a Lizmap JSON configuration file for setting up the Lizmap parameters. Data inserted, modified or deleted during the lifecycle of the platform are immediately available (or hidden in the case of deletion) on the \emph{Map} as soon as the operation is completed.

Figure \ref{fig:demo} shows four examples of interactions with the \emph{map}. In \ref{fig:demo} (a) it is possible to see the base map brought by OpenStreetMap~\cite{OpenStreetMap}; in \ref{fig:demo} (b) the user has selected the visualization of a raster layer;  in \ref{fig:demo} (d) the user zoomed on a field for observing the acquisitions.

\newcommand{\icon}[1]{{\includegraphics[width=0.6cm]{#1}}}
\newcommand{\iconrow}[3]{
    \icon{#1} & \textbf{#2}: #3 \\
    \vspace{.1cm}
}

\begin{table}
\centering
\resizebox{0.8\textwidth}{!}{
\begin{tabularx}{\textwidth}{ m{0.5cm} m{7cm} }
    \iconrow{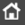}{Home}{brings the user back to the project selection page.}
    \iconrow{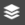}{Layers}{to control the visibility of the layers on the map}
    \iconrow{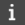}{Information}{gives information about the project}
    \iconrow{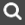}{Filter}{lets the user filter data based on their properties.}
    \iconrow{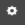}{WPS}{for choosing a model (or algorithm) uploaded and extract useful informations. }
    \iconrow{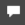}{Popup}{it will show information about the data clicked by the user.}
    \iconrow{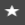}{Selection}{enables the drawing on the map.}
    \iconrow{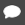}{Tooltip}{It will highlight data of the layer under inspection that lies under the cursor.}
    \iconrow{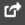}{Permalink}{enables the sharing of the link.}
    \iconrow{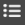}{Data}{shows informations about collected data in a tabular way.}
    \iconrow{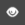}{WPSResults}{shows the results of the algorithms used.}

\end{tabularx}
}
\vspace{-0.4cm}
\caption{Utilities included in the web-map.}
\label{tab:lizmap-utilities}
\end{table}

\subsection{Tabular}
\label{sub:backend}

\begin{figure*}[!ht]
\begin{center}
\includegraphics[width=.7\linewidth]{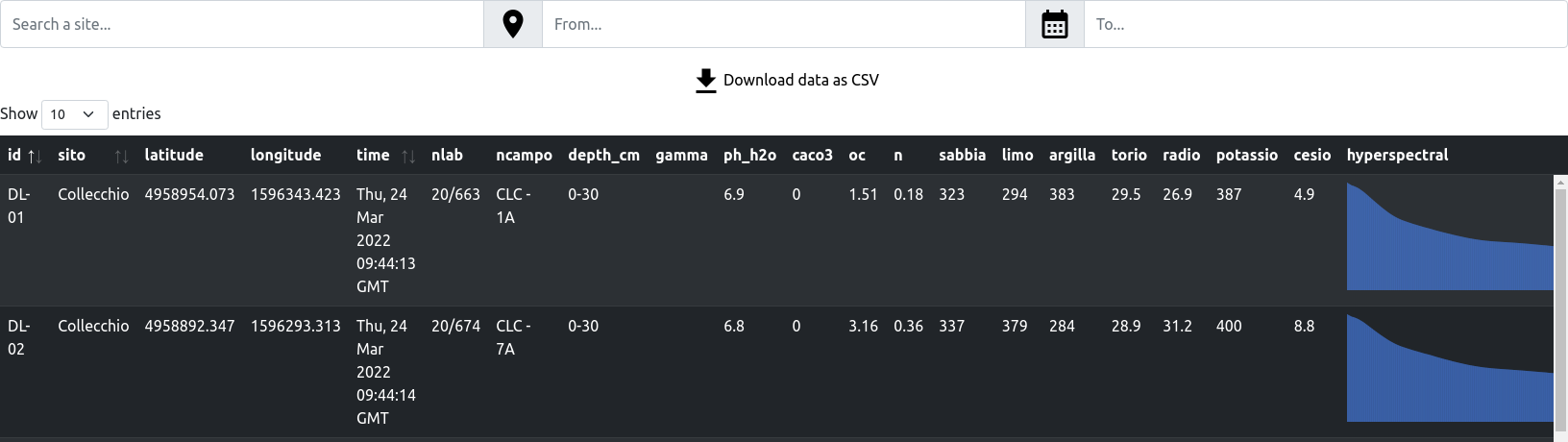} \\
\end{center}
\caption{Closeup of the \emph{Tabular} interface. Each sample is identified through an id and spatially located by its GPS coordinates. Each sample is characterized by its soil properties and by its hyperspectral signal, here reported graphically for a better user experience.}
\label{fig:closeups}
\end{figure*}

The \emph{Tabular} displays data in tabular format. It is composed by two pages for handling respectively the acquisitions $\mathcal{L}$ and $\mathcal{D}$. Figure \ref{fig:closeups} shows an excerpt of $\mathcal{L}$ samples. In this context, each lab sample is characterized by a numerical id, the position expressed in latitude and longitude and all the chemical properties related to it. To make the visualization of the  samples more effective, the last column includes a graphical plot of the hyperspectral signal instead of the raw values.

While the \emph{map} is visible without authentication, the \emph{Tabular} is accessible only to registered users. Among them, two types are identified: the \textit{normal users}, who can view, filter, sort and download data and the \textit{administrators} who can go further and modify the data stored in the dataset. In particular, they can upload one ore more samples of $\mathcal{L}$ and $\mathcal{D}$ at once through a CSV file, upload zipped archive of images or new predictive models that can be later used for estimating variables. 

The \emph{Tabular} is a Flask-based web application developed in Python that queries the data through SQLAlchemy. Information are gathered through the browser and thus rendered with HTML/CSS/JS. Two external libraries are used: Datatable\footnote{\url{https://datatables.net/}} for rendering the tables and Peity\footnote{\url{http://benpickles.github.io/peity/}} for plotting the spectral signals.

\subsection{API}
\label{sub:api}

The platform \emph{API} allows the uploading, querying and editing of $\mathcal{D}$ samples. New samples can be loaded in real-time from a software running on the acquisition device. Two modalities have been identified: a single and a massive upload. The former can be performed using the HTTP protocol and a JSON-formatted request while the latter through a CSV file.

For convenience, the \emph{API} is included in the same Flask web app of the \emph{Tabular} as the two elements share several functions. It has been implemented following a REST (Representational state transfer) style whose principle nicely fits with the HTTP protocol used for communication.

The JSON Web Token (JWT) open standard is used to improve the security of the system. In first place the user authenticates into the system and receives, through an HTTP response, a token that represents its identity. Further on, this token enables the use of \emph{API} capabilities.

\begin{figure}[!ht]
\centering
\includegraphics[width=0.9\linewidth]{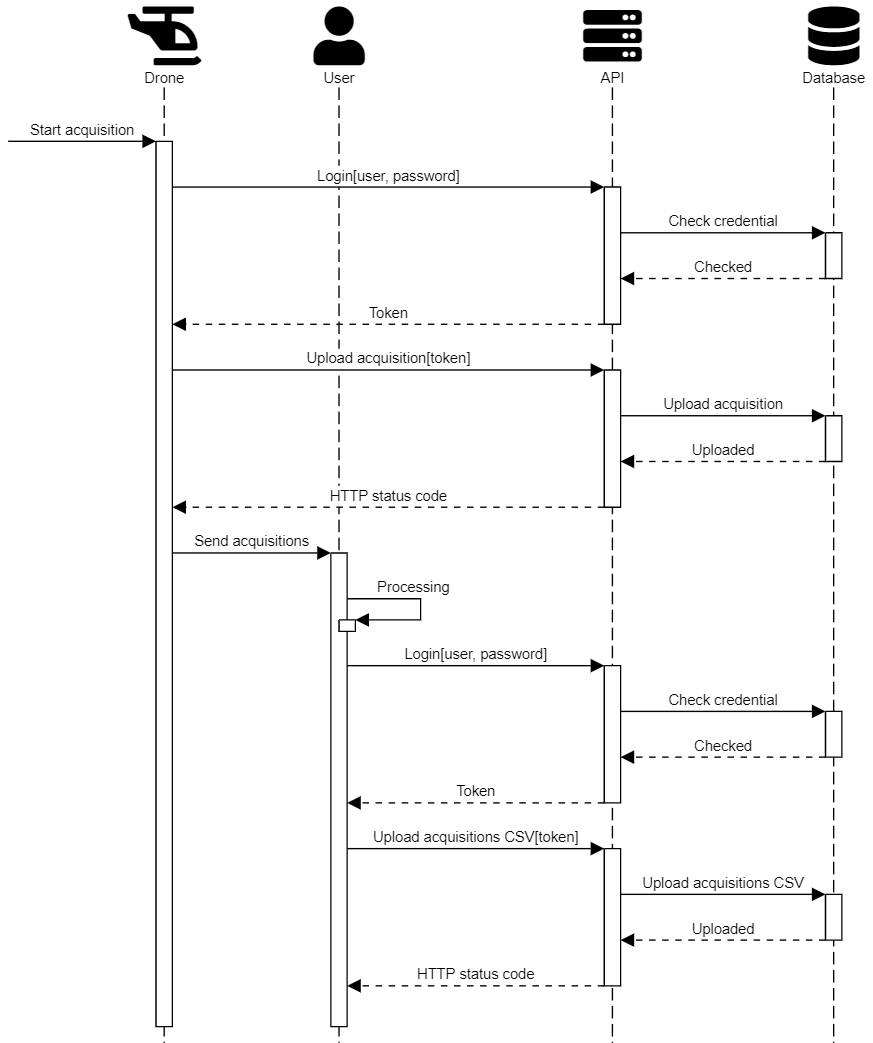}
\caption{High-level flow of API's execution. On-board devices and users can interact programmatically with the system through the \emph{API}. After the logging procedure performed through token, single or massive upload of acquisitions can be performed.}
\label{fig:api}
\end{figure}

An example of the API usage is presented in Fig.\ref{fig:api}. In this example, we show the communication flow between the user, or an acquisition device, and the platform. In first place, the user  requires the authorization to interact with the database through the \emph{middle layer}, and in second place it performs a single and massive data uploading.

\subsection{Database}
\label{sub:db}

\begin{figure}[tb]
\centering
\begin{adjustbox}{width=.99\columnwidth}
\includegraphics[width=\columnwidth]{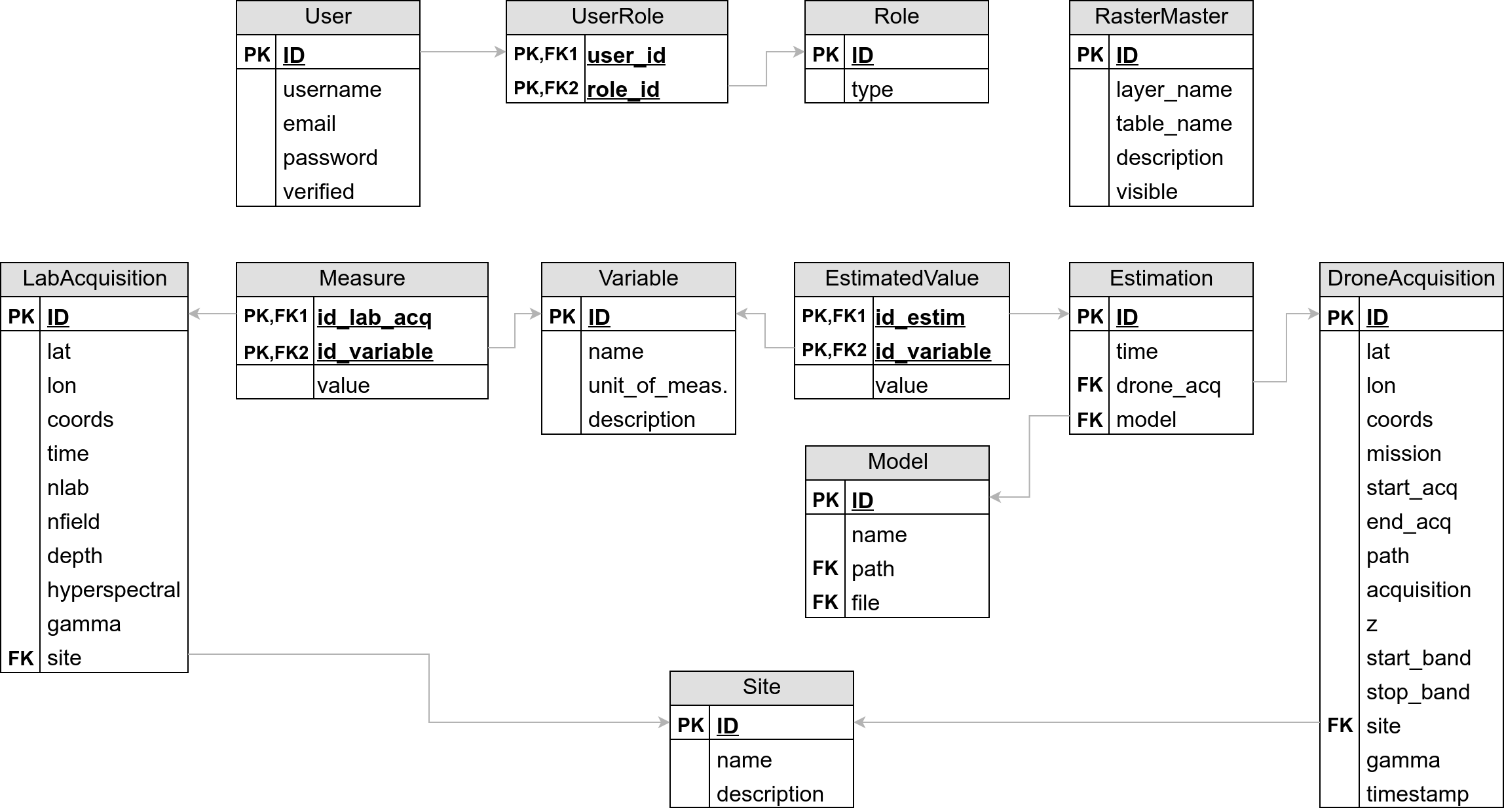}
\end{adjustbox}
\caption{Database organization. It considers five aspects, the handling of users, the acquisitions made in laboratory and with airborne sensors, the rasters and finally the predictions made by predictive models on airborne acquired data. The considered variables are not a static set but they can easily change as the researchers focus on different characteristics of soil.}
\label{fig:db-er}
\end{figure}

The \emph{database} is designed to handle registered users, laboratory ($\mathcal{L}$) and drone ($\mathcal{D}$) acquisitions as well as machine learning predictions and remote sensing rasters. Figure \ref{fig:db-er} shows the logical organization of the platform database with the main entities and relationships between them. 

Let $s_L$ and $s_D$ be respectively two samples of $\mathcal{L}$ and $\mathcal{D}$. $s_L$ and $s_D$ are located by means of geographic coordinates (latitude and longitude). They can belong to zero as well as multiple \textit{sites}. A \textit{site} is a logical grouping of acquisitions, such as a sampling campaign or a field. 

Each $s_L$ consists of a hyperspectral signal associated with the measurement of one or more soil variables. This flexibility allows the set of soil properties considered to be modified at any time. For example, separate acquisition campaigns considering different sets of variables can be modeled seamlessly in the system. These samples constitute the set of data used for the training of supervised predictors.

Samples $s_D$ are not associated with soil properties. In fact, they represent specimens loaded at test time that require prediction by a predictive model. For this purpose, they are associated in the database with one or more estimates that in turn are associated with the model that made them, in combination with the variables that were predicted by that model. This allows different variables to be predicted by multiple predictive models.

Each raster is represented through a piramidal view in order to decrease the loading time. Each raster is therefore represented by a meta table that contains the information of each layer of the pyramid associated to the proper level of zoom. We also designed the table \textit{RasterMaster} to keep track of all the rasters in the system, with the possibility to disable the visualization of a raster at runtime. %

The platform \emph{database} has been developed in PostgreSQL. To support georeferenced data, we enriched the suite adopting PostGIS. This extension allows the use of geographic objects following the specifications from the Open Geospatial Consortium (OGC)~\footnote{\url{https://www.ogc.org/}}.

\section{Deployment}
\label{sec:deployment}

The server is hosted on an Ubuntu 20.04 machine composed by an ASRock Z68 Extreme3 Gen3 motherboard and an Intel(R) Core(TM) i5-2500K CPU @ 3.30GHz. All code is included in a Docker image containing all the dependencies. The web server used is NGINX.
A DNS record has been binded to the server ip to make the platform easily accessible through the URL  \url{http://pignoletto.lombardia.it}.

\section{Usability}
\label{sec:usability}

\begin{figure}
\centering
\begin{tabularx}{\textwidth}{cc}
\includegraphics[width=.45\columnwidth]{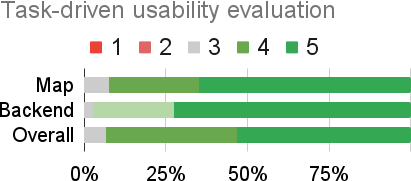} & 
\includegraphics[width=.45\columnwidth]{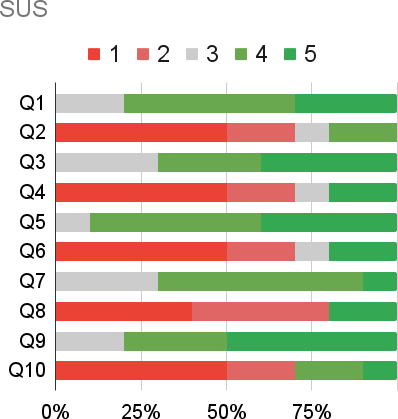} \\
(a) & (b) \\
\end{tabularx}
\caption{Usability results for (a) task-driven (TDS) and (b) SUS assessments. For TDS and odd questions of SUS, the higher the better. For even questions of SUS, the lower the better.}
\label{fig:usability-results}
\end{figure}

Two usability tests were performed to evaluate the ease of use of the \emph{Map} and the \emph{Tabular} components. The first one is a task-driven survey (TDS) containing the 11 questions showed in table \ref{tab:task-driven-questions}. Each question asks to perform a task and assess the easiness of doing it. In this case, the higher is the score, the better is the goodness.
The second one is the System Usability Scale (SUS) \cite{brooke1996sus}. For odd questions, the higher the better. At the contrary, for even questions, the lower the better. 10 users were involved in the usability assessment. Figures \ref{fig:usability-results} (a) and (b) show respectively the results of the TDS and the SUS surveys. Both evaluations demonstrated that the \emph{Map} and the \emph{Tabular}  are easy to use and the learning curve is low. From the analysis of the replies to question 3 of TDS, it emerged that users prefer more levels of zoom, which has been corrected after the evaluation.

\section{Conclusions}
\label{sec:conclusion}

In this paper we presented to the state of art a new platform for handling and processing airborne and proximity data. It also capable of integrating predictive models and their relative predictions.
The presented tool intends to support all stakeholders related to the agricoltural sector in adopting the best practices of PA with the aim of improving the entire agricultural production chain.
Although it is initially published to support the specific needs of the Lombardy region, the platform is easily extendable to other locations with different needs, thus making it a software exploitable for any purpose.
The usability assessment that we performed in section \ref{sec:usability} showed that the presented system is easy to use and the learning curve is low.
Those factors, in combination with the technical gain offered by the platform, suggest that the presented system can bring a noticeable impact on the agricoltural sector thanks to its effective utility and ease of use.
Thus, supporting the transition to PA techniques to make agriculture more efficient, sustainable and less impacting on the environment and climate.

\section*{Acknowledgement}
Research developed in the context of the project PIGNOLETTO $-$ Call HUB Ricerca e Innovazione CUP (Unique Project Code) n. E41B20000050007, co-funded by POR FESR 2014-2020 (Programma Operativo Regionale, Fondo Europeo di Sviluppo Regionale – Regional Operational Programme, European Regional Development Fund).

\section*{Conflict of Interest}
\label{sec:COI}
No conflict of interest exists: We wish to confirm that there are no known conflicts of interest associated with this publication and there has been no significant financial support for this work that could have influenced its outcome.

\printcredits

\bibliographystyle{elsarticle-num}
\bibliography{cas-refs}

\begin{thebibliography}{10}
\expandafter\ifx\csname url\endcsname\relax
  \def\url#1{\texttt{#1}}\fi
\expandafter\ifx\csname urlprefix\endcsname\relax\def\urlprefix{URL }\fi
\expandafter\ifx\csname href\endcsname\relax
  \def\href#1#2{#2} \def\path#1{#1}\fi

\bibitem{velten2015sustainable}
S.~Velten, J.~Leventon, N.~Jager, J.~Newig, What is sustainable agriculture? a
  systematic review, Sustainability 7~(6) (2015) 7833--7865.

\bibitem{pierce1999aspects}
F.~J. Pierce, P.~Nowak, Aspects of precision agriculture, Advances in agronomy
  67 (1999) 1--85.

\bibitem{lindblom2017promoting}
J.~Lindblom, C.~Lundstr{\"o}m, M.~Ljung, A.~Jonsson, Promoting sustainable
  intensification in precision agriculture: review of decision support systems
  development and strategies, Precision Agriculture 18~(3) (2017) 309--331.

\bibitem{shafi2019precision}
U.~Shafi, R.~Mumtaz, J.~Garc{\'\i}a-Nieto, S.~A. Hassan, S.~A.~R. Zaidi,
  N.~Iqbal, Precision agriculture techniques and practices: From considerations
  to applications, Sensors 19~(17) (2019) 3796.

\bibitem{ayaz2019internet}
M.~Ayaz, M.~Ammad-Uddin, Z.~Sharif, A.~Mansour, E.-H.~M. Aggoune,
  Internet-of-things (iot)-based smart agriculture: Toward making the fields
  talk, IEEE access 7 (2019) 129551--129583.

\bibitem{sishodia2020applications}
R.~P. Sishodia, R.~L. Ray, S.~K. Singh, Applications of remote sensing in
  precision agriculture: A review, Remote Sensing 12~(19) (2020) 3136.

\bibitem{tantalaki2019data}
N.~Tantalaki, S.~Souravlas, M.~Roumeliotis, Data-driven decision making in
  precision agriculture: the rise of big data in agricultural systems, Journal
  of Agricultural \& Food Information 20~(4) (2019) 344--380.

\bibitem{linaza2021data}
M.~T. Linaza, J.~Posada, J.~Bund, P.~Eisert, M.~Quartulli, J.~D{\"o}llner,
  A.~Pagani, I.~G~Olaizola, A.~Barriguinha, T.~Moysiadis, et~al., Data-driven
  artificial intelligence applications for sustainable precision agriculture,
  Agronomy 11~(6) (2021) 1227.

\bibitem{mulla2021satellite}
D.~J. Mulla, Satellite remote sensing for precision agriculture, in: Sensing
  Approaches for Precision Agriculture, Springer, 2021, pp. 19--57.

\bibitem{elijah2018overview}
O.~Elijah, T.~A. Rahman, I.~Orikumhi, C.~Y. Leow, M.~N. Hindia, An overview of
  internet of things (iot) and data analytics in agriculture: Benefits and
  challenges, IEEE Internet of things Journal 5~(5) (2018) 3758--3773.

\bibitem{strickland2022andrew}
E.~Strickland, Andrew ng, ai minimalist: The machine-learning pioneer says
  small is the new big, IEEE Spectrum 59~(4) (2022) 22--50.

\bibitem{kukar2019agrodss}
M.~Kukar, P.~Vra{\v{c}}ar, D.~Ko{\v{s}}ir, D.~Pevec, Z.~Bosni{\'c}, et~al.,
  Agrodss: A decision support system for agriculture and farming, Computers and
  Electronics in Agriculture 161 (2019) 260--271.

\bibitem{coppola2019landsupport}
E.~Coppola, F.~D. Moccia, Landsupport, a decision support system for
  territorial government, UPLanD-Journal of Urban Planning, Landscape \&
  environmental Design 4~(2) (2019) 29--38.

\bibitem{longley2005geographic}
P.~A. Longley, M.~F. Goodchild, D.~J. Maguire, D.~W. Rhind, Geographic
  information systems and science, John Wiley \& Sons, 2005.

\bibitem{QGIS}
{QGIS Development Team}, \href{https://www.qgis.org}{QGIS Geographic
  Information System}, QGIS Association (2022).
\newline\urlprefix\url{https://www.qgis.org}

\bibitem{booth2001getting}
B.~Booth, A.~Mitchell, et~al., Getting started with arcgis (2001).

\bibitem{sebastianelli2021automatic}
A.~Sebastianelli, M.~P. Del~Rosso, S.~L. Ullo, Automatic dataset builder for
  machine learning applications to satellite imagery, SoftwareX 15 (2021)
  100739.

\bibitem{ma2001mapping}
Z.~Ma, M.~M. Hart, R.~L. Redmond, Mapping vegetation across large geographic
  areas: integration of remote sensing and gis to classify multisource data,
  Photogrammetric Engineering and Remote Sensing 67~(3) (2001) 295--308.

\bibitem{frigerio2016gis}
I.~Frigerio, S.~Ventura, D.~Strigaro, M.~Mattavelli, M.~De~Amicis, S.~Mugnano,
  M.~Boffi, A gis-based approach to identify the spatial variability of social
  vulnerability to seismic hazard in italy, Applied geography 74 (2016) 12--22.

\bibitem{blanco2018agricultural}
I.~Blanco, R.~V. Loisi, C.~Sica, E.~Schettini, G.~Vox, Agricultural plastic
  waste mapping using gis. a case study in italy, Resources, Conservation and
  Recycling 137 (2018) 229--242.

\bibitem{mancini2010gis}
F.~Mancini, C.~Ceppi, G.~Ritrovato, Gis and statistical analysis for landslide
  susceptibility mapping in the daunia area, italy, Natural Hazards and Earth
  System Sciences 10~(9) (2010) 1851--1864.

\bibitem{ladisa2012gis}
G.~Ladisa, M.~Todorovic, G.~T. Liuzzi, A gis-based approach for desertification
  risk assessment in apulia region, se italy, Physics and Chemistry of the
  Earth, Parts a/B/C 49 (2012) 103--113.

\bibitem{michelon2019software}
G.~K. Michelon, C.~L. Bazzi, S.~Upadhyaya, E.~G. de~Souza, P.~S.~G.
  Magalh{\~a}es, L.~F. Borges, K.~Schenatto, R.~Sobjak, A.~Gavioli, N.~M.
  Betzek, Software agdatabox-map to precision agriculture management, SoftwareX
  10 (2019) 100320.

\bibitem{bazzi2019agdatabox}
C.~L. Bazzi, E.~P. Jasse, P.~S.~G. Magalh{\~a}es, G.~K. Michelon, E.~G.
  de~Souza, K.~Schenatto, R.~Sobjak, Agdatabox api--integration of data and
  software in precision agriculture, SoftwareX 10 (2019) 100327.

\bibitem{petito2022impact}
M.~Petito, S.~Cantalamessa, G.~Pagnani, F.~Degiorgio, B.~Parisse, M.~Pisante,
  Impact of conservation agriculture on soil erosion in the annual cropland of
  the apulia region (southern italy) based on the rusle-gis-gee framework,
  Agronomy 12~(2) (2022) 281.

\bibitem{laurent2018online}
J.-B. Laurent, L.~Leroux, Online publication of a land cover map using lizmap,
  QGIS and Generic Tools 1 (2018) 243--255.

\bibitem{michaelis2008web}
C.~D. Michaelis, D.~P. Ames, Web feature service (wfs) and web map service
  (wms). (2008).

\bibitem{anderson2015docker}
C.~Anderson, Docker [software engineering], Ieee Software 32~(3) (2015)
  102--c3.

\bibitem{OpenStreetMap}
{OpenStreetMap contributors}, {Planet dump retrieved from
  https://planet.osm.org }, \url{ https://www.openstreetmap.org } (2017).

\bibitem{brooke1996sus}
J.~Brooke, Sus: a “quick and dirty’usability, Usability evaluation in
  industry 189~(3).

\end{thebibliography}

\newpage
\section*{Appendix}
\subsection*{Code Metadata}

Table \ref{tab:code-metadata} contains the metadata of the code described in this paper, with all the information necessary for the download.

\begin{table*}[H]
\begin{tabular}{l l}
\toprule
Current code version & 1.0 \\
Permanent link to code/repository used for this code version & \href{https://github.com/SimoLoca/Pignoletto\_platform}{https://github.com/SimoLoca/Pignoletto\_platform} \\
Legal Code License   & MIT \\
Code versioning system used & Git \\
Software code languages, tools, and services used & Python, Javascript, CSS, HTML, PHP  \\
Compilation requirements, operating environments \& dependencies & Docker \\
If available Link to developer documentation/manual & \href{https://github.com/SimoLoca/Pignoletto\_platform/wiki}{https://github.com/SimoLoca/Pignoletto\_platform/wiki} \\
Support email for questions & s.locatelli29@campus.unimib.it \\
\bottomrule
\end{tabular}
\caption{Code metadata description}
\label{tab:code-metadata}
\end{table*}

\subsection*{Questions of task-driven assessment}
\label{sec:questions}

Table \ref{sec:questions} contains the questions posed to the users in the TDS test.

\begin{table*}
\renewcommand{\arraystretch}{1.4}
\centering
\begin{tabularx}{\textwidth}{l m{16cm}}
\toprule

\multirow{4}{*}{\rotatebox[origin=c]{90}{\parbox[c]{30mm}{\centering \textbf{Map}}}}
& Display the acquisitions relative to the Argilla variable under the group lab\_acquisition. Zoom on a point and click on it to view its value. How hard is this task? \\

& Open the tabular visualization ("Dati" tool on the left sidebar) and click on the button relative to the Argilla acquisitions to open the table. Locate the sample DL-06 in the table and click the lens icon to view the corresponding point on the map. Click on the found point to view the relative info on the left panel \\

& Zoom out to have all the acquisitions (points) on the screen visible. Search all the samples that contain a level of Argilla between 200 and 400 (use the "Filtro" tool on the left sidebar) \\

& Display the raster slope (it can take some time to load) \\

\midrule

\multirow{4}{*}{\rotatebox[origin=c]{90}{\parbox[c]{25mm}{\centering \textbf{Tabular}}}}
& Download the dataset about Laboratory acquisition \\
& Filter the samples that have been acquired in Lodi \\
& Sort the "sito" column on the Laboratory acquisitions in descending order \\
& Go to the Drone acquisition page, try to filter the samples that have been acquired from April 5th 2022 to May 14th 2022 \\

\midrule
\multirow{3}{*}{\rotatebox[origin=c]{90}{\parbox[c]{15mm}{\centering \textbf{General}}}}
& How user-friendly is Pignoletto platform map interface? \\
& How user-friendly is Pignoletto platform tabular interface? \\
& Do you think this application can provide valuable support for precision agriculture? \\
\bottomrule
\end{tabularx}
\caption{Task driven questions}
\label{tab:task-driven-questions}
\end{table*}

\end{document}